\documentclass[Physsubmission, Phys]{SciPost}

\hypersetup{
    colorlinks,
    linkcolor={red!50!black},
    citecolor={blue!50!black},
    urlcolor={blue!80!black}
}

\usepackage[bitstream-charter]{mathdesign}
\usepackage{mcite}
\DeclareSymbolFont{usualmathcal}{OMS}{cmsy}{m}{n}
\DeclareSymbolFontAlphabet{\mathcal}{usualmathcal}

\begin{document}

\begin{center}{\Large \textbf{
Study of $\text{D}^0$ meson interactions via femtoscopic correlations\\
}}\end{center}

\begin{center}
Isabela M. Silvério\textsuperscript{1$\star$},
Sandra S. Padula\textsuperscript{1} and
Gastão I. Krein\textsuperscript{1}
\end{center}

\begin{center}
{\bf 1} Instituto de Física Teórica, Universidade Estadual Paulista
\\
*isabela.maietto@gmail.com
\end{center}

\begin{center}
\today
\end{center}


\definecolor{palegray}{gray}{0.95}
\begin{center}
\colorbox{palegray}{
  \begin{tabular}{rr}
  \begin{minipage}{0.1\textwidth}
    \includegraphics[width=30mm]{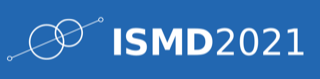}
  \end{minipage}
  &
  \begin{minipage}{0.75\textwidth}
    \begin{center}
    {\it 50th International Symposium on Multiparticle Dynamics}\\ {\it (ISMD2021)}\\
    {\it 12-16 July 2021} \\
    \doi{10.21468/SciPostPhysProc.?}\\
    \end{center}
  \end{minipage}
\end{tabular}
}
\end{center}

\section*{Abstract}
{\bf
Femtoscopy is a powerful tool that can be used to investigate the space-time dimensions of the region from which the particles are emitted. When applied to high energy collisions this method is sensitive not only to quantum statistics, but also to final state interactions, such as strong interactions between hadrons and Coulomb interactions when considering charged particles. In particular, this procedure is used to investigate the sensitivity of the strong interaction to different source sizes for $\textrm{D}^{0}$ mesons correlations.}
\section{Introduction}
\label{sec:intro}
The studies of identical particles correlations started in the mid-fifties, with Robert Hanbury-Brown and Richard Q. Twiss performing a measurement of the angular diameter of a star using the correlation between two photons emitted from it \cite{brown1954}. This leads to the intensity interferometry or Hanbury-Brown-Twiss (HBT) effect. Later in 1960, G. Goldhaber, S. Goldhaber, W. Y. Lee and A. Pais \cite{GGLP60} performed an experiment intended to discover the $\rho$ meson by measuring charged pions in proton-antiproton collisions at $\sqrt{s}=1.05$ GeV. The study showed that the momentum correlation of two identical pions from the same event could be described as the Fourier transform of a function of the phase space distribution of the emission source. However, it was later demonstrated that the method applied to high energy collisions is sensitive not only to quantum statistics (QS) \cite{lisa2005}, but also to final state interactions, such as Coulomb and strong interactions (SI). In this work, the construction of the correlation function will be discussed, considering the description for QS and final state interactions contributions, focusing mainly on SI, since the pair of particles of interest are electrically neutral. In particular, this procedure is used to investigate the sensitivity of the strong interaction to different source sizes for $\text{D}^{0}\text{D}^{0}$, $\overline{\text{D}}^{\hspace{0.05cm}0}\overline{\text{D}}^{\hspace{0.05cm}0}$ correlations, which can contribute for heavy particle correlations studies and to investigate new exotic bound states of matter.

\section{Two-particle correlation function}
\label{cf}
The two-particle correlation function (CF) can be approximated as \cite{lisa2005}
\begin{equation}
C\left(\mathbf{k}_{1},\mathbf{k}_{2}\right)=\frac{P\left(\mathbf{k}_{1},\mathbf{k}_{2}\right)}{P\left(\mathbf{k}_{1}\right)P\left(\mathbf{k}_{2}\right)}\Rightarrow C(k)\approx\int{d^{3}r\;S(r)|\Psi(\mathbf{r},\mathbf{k})|^{2}},
\end{equation}
where $\Psi(\mathbf{r},\mathbf{k})$ is the total wave function, and $S(r)$ is assumed to be a Gaussian source function $S(r)=\frac{\text{e}^{-r^{2}/4R^{2}}}{(4\pi R^{2})^{3/2}}$, with $R$ being the width of the source. In the case of two identical bosons, assuming the presence of a potential that only affects the s-wave, the two-particle wave function can be written as
\begin{equation}
\Psi(\mathbf{r},\mathbf{k})=\sqrt{2}\left[\cos{\left(\mathbf{k}\cdot\mathbf{r}\right)}-j_{0}(kr)+\psi_{0}(r)\right],
\end{equation}
where $\mathbf{r}=\mathbf{r_{1}}-\mathbf{r_{2}}$, $\mathbf{k}=\left(\mathbf{k_{1}}-\mathbf{k_{2}}\right)/2$, $j_{0}(kr)$ is the spherical Bessel function, and $\psi_{0}(r)$ is the scattered wave function for $l=0$.
Finally, the correlation function between two identical bosons is written as
\begin{equation}
C(k)=1+e^{-4k^{2}R^{2}}+2\int{d^{3}r\;S(r)\left[|\psi_{0}(r)|^{2}-j^{2}_{0}(kr)\right]}.
\end{equation}
Considering the assymptotic form for $\psi_{0}(r)$, the Lednicky Lyuboshitz (LL) model \cite{Lednicky:1981su} is obtained, giving an approximated description of the strong interactions effects felt by two identical bosons, as
\begin{equation}
\label{LLmodel}
    C_{\textrm{LL}}(k)=1+2\left[\frac{|f_{0}(k)|^{2}}{2R^{2}}\left(1-\frac{r_{0}}{2\sqrt{\pi}R}\right)+\frac{2\mathrm{Re}f_{0}(k)}{\sqrt{\pi}R}F_{1}(2kR)-\frac{\mathrm{Im}f_{0}(k)}{R}F_{2}(2kR)\right].
\end{equation}
In the expression above the effective range expansion (ERE) is considered, i.e.,
\begin{equation}
\label{ERE}
f_{0}(k)=\left[k\cot{\delta_{0}(k)}-ik\right]^{-1}\approx  \left[-\frac{1}{a_{0}}+\frac{1}{2}r_{0}k^{2}-ik\right]^{-1},
\end{equation}
where $\delta_{0}$ is the phase shift, $a_{0}$ is the scattering length and $r_{0}$ is the effective range.

\subsection{Numerical description}
\label{sec:another}
To obtain the correlation function, the total wave function of the system is needed. In this study, a numerical description is employed to solve the Lippmann-Schwinger (LS) equation, for the case of non-local potentials, which is the case for $\text{D}^{0}\text{D}^{0}$, $\overline{\text{D}}^{\hspace{0.05cm}0}\overline{\text{D}}^{\hspace{0.05cm}0}$ correlations \cite{Hadjimichef:1998rx}. 
The LS equation is given by
\begin{equation}
\hat{T}(k^{2})=\hat{V}+\hat{V}\hat{G}(k^{2})\hat{T}(k^{2}).
\end{equation}
where $\hat{T}$ is the transition operator, $\hat{V}$ is the potential matrix, and $\hat{G}$ is the Green function. The scattered wave function can be written in terms of the $\hat{T}$ for the s-wave, from the LS equation, as
\begin{equation}
\psi_{0}(k,r)=j_{0}(kr)+\frac{2}{\pi}\int^{\infty}_{0}dq q^{2}\frac{j_{0}(qr)T_{0}(q,k;k^{2})}{(k^{2}-q^{2})/2\mu\pm i\epsilon}.
\end{equation}


\section{Results and Summary}

\begin{figure}[h]
\centering
\includegraphics[width=0.3\linewidth]{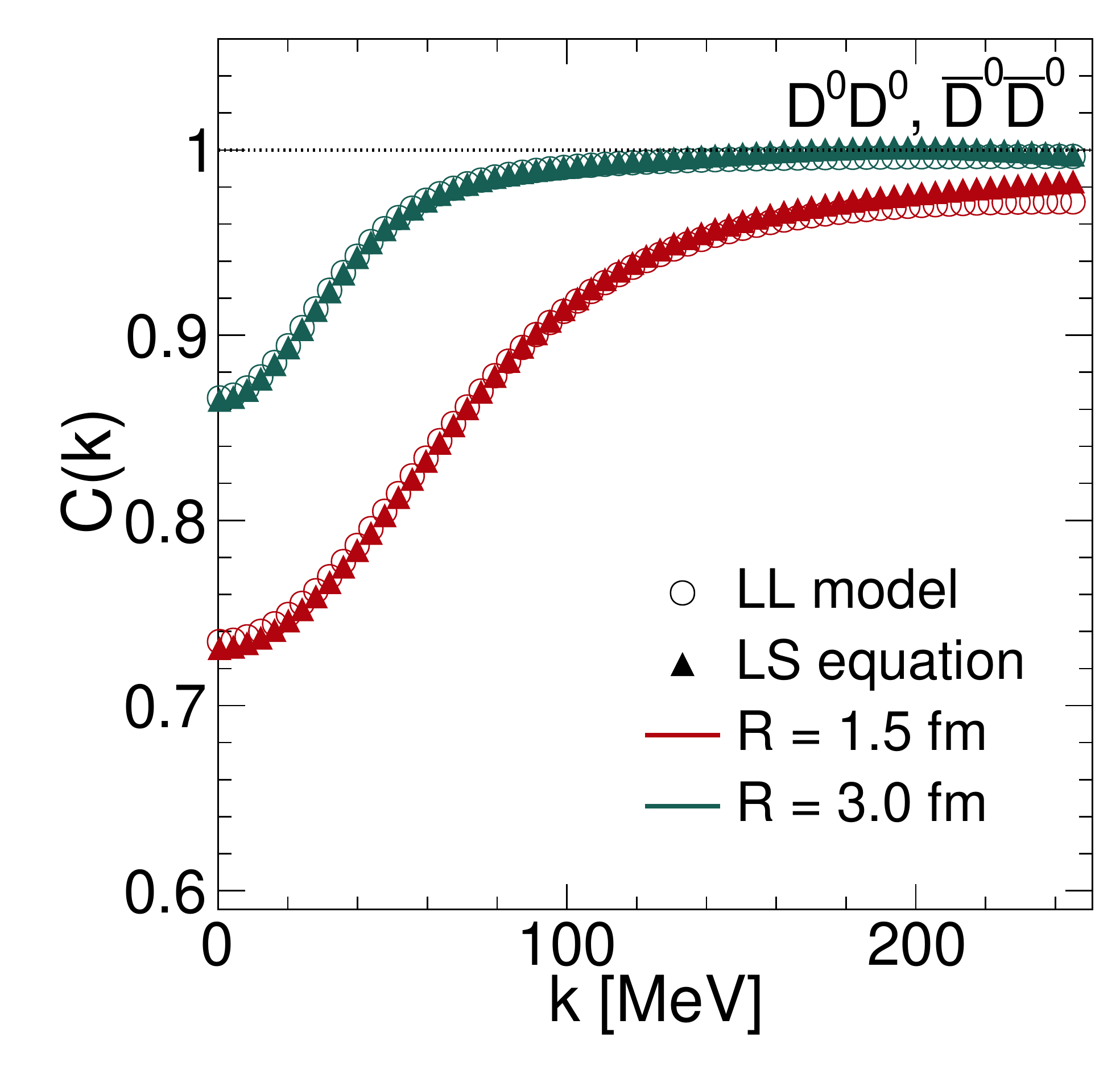}\includegraphics[width=0.3\linewidth]{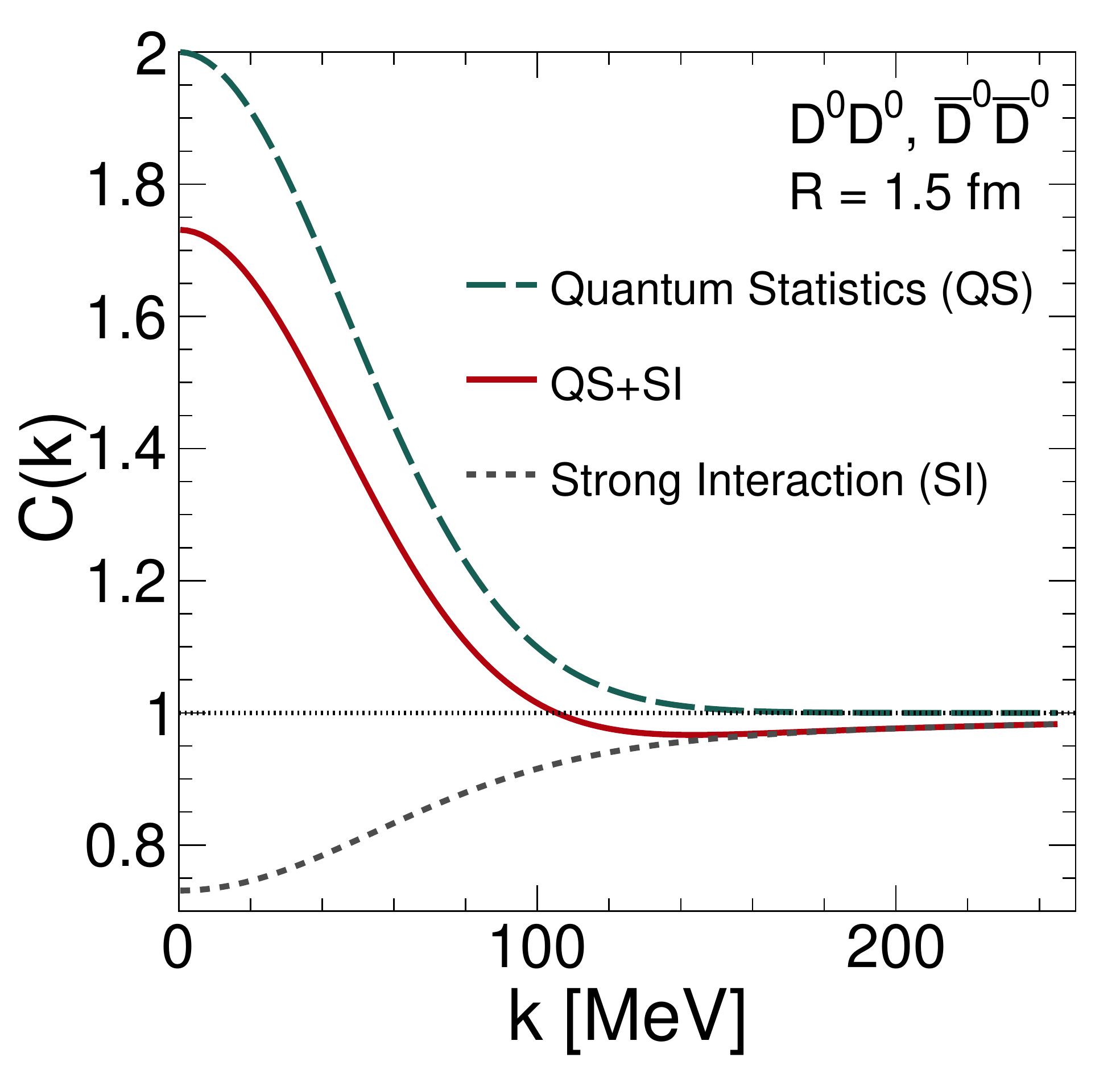}\includegraphics[width=0.3\linewidth]{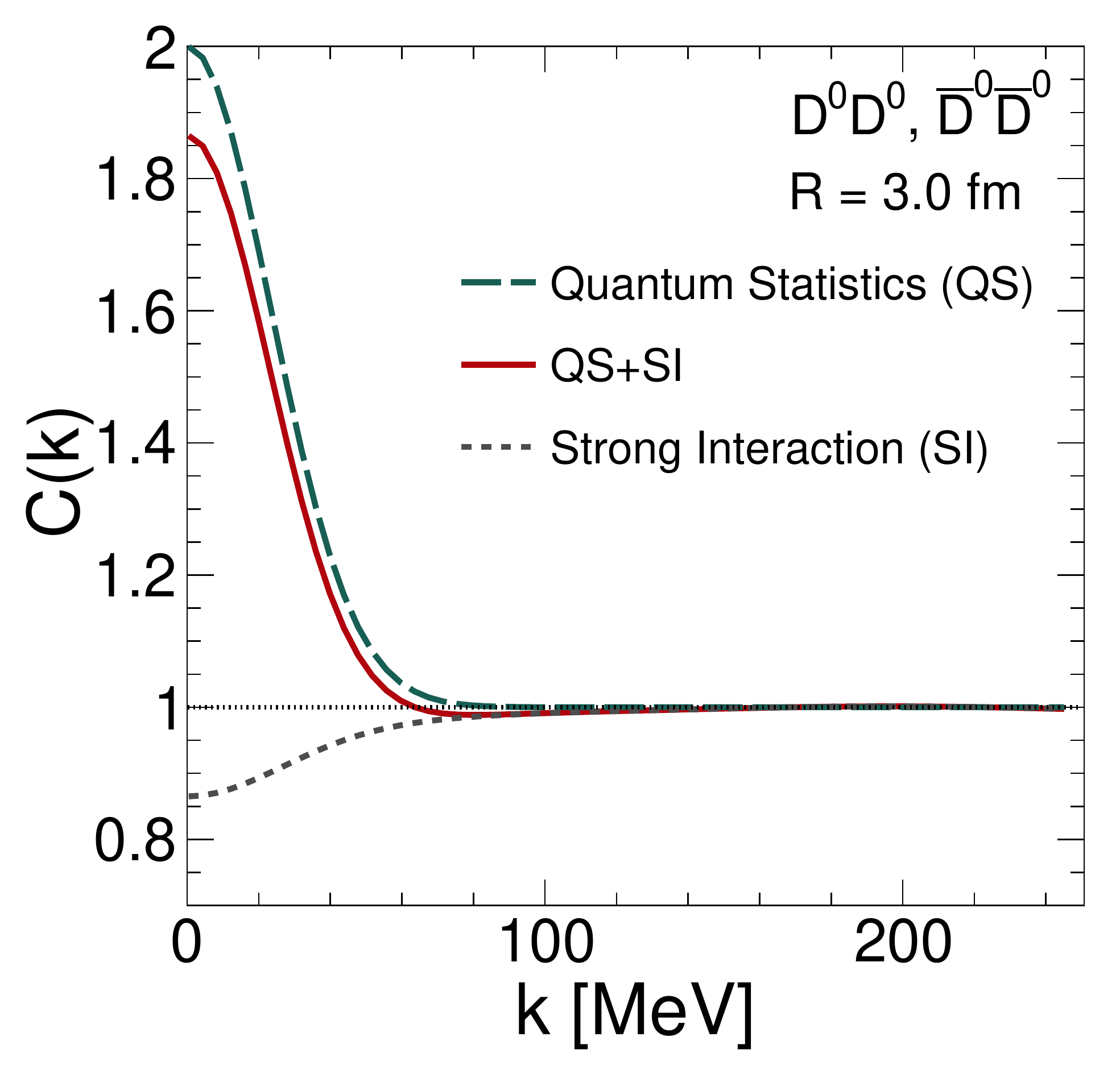}
\caption{\textbf{Left}: LL model (empty circles) and CF obtained from LS equation code (filled triangles) for two values of $R$: $1.5$ fm (red) and $3.0$ fm (green). \textbf{Center, Right}: Two-particle CF contributions: SI (grey, short dashed line), total (QS+SI; red, full line) and QS (green, long dashed line) as a function of momentum $k$, using the source width as $R=1.5\;\textrm{fm}$ (left) and $R=3.0\;\textrm{fm}$ (right).}
\label{CF}
\end{figure}

From the numerical method the scattering observables were extracted, implemented in Eq. (\ref{LLmodel}) and the result was compared with the one obtained with the code, Fig. \ref{CF} (left). The code provided the $\delta_{0}$ used to fit Eq. (\ref{ERE}) and obtain $a_{0}=0.182$ fm and $r_{0}=2.44$ fm.
Finally, it is possible to obtain the curves for the correlation function contributions, such as the QS and SI contributions for $R=1.5$ fm and $R=3.0$ fm, Fig. \ref{CF} (center, right).

\section{Conclusion}
The results obtained for  $\textrm{D}^{0}\textrm{D}^{0}$, $\overline{\textrm{D}}^{\hspace{0.05cm}0}\overline{\textrm{D}}^{\hspace{0.05cm}0}$ correlations show that the QS contribution reproduces the behavior of QS correlations between identical spin-zero bosons, since it is expected that the correlation function goes above one, which in femtoscopy is called an "attractive" correlation. It was shown that the effect of the strong interaction is more significant at smaller values of R, since the probability of emission of the bosons from closer points is larger and more likely the relative distance of the pair is smaller, where the short-range contribution of the potential predominates. Finally, with this study, it was possible to show that femtoscopy can be used to study final-state interactions and give information about the shape of the particle correlation functions. Furthermore, the results in this work might be used for correlation studies of  $\textrm{D}^{0}\textrm{D}^{0}$ and $\overline{\textrm{D}}^{\hspace{0.05cm}0}\overline{\textrm{D}}^{\hspace{0.05cm}0}$ in particle collider experiments, as a guideline.

\section*{Acknowledgements}
This study was financed in part by: Coordena\c{c}\~ao Aperfei\c{c}oamento 
de Pessoal de N\'{\i}vel Superior - CAPES, grant no. 88887-357054/2019-00 (I.M.S.), 
Conselho Nacional de Densenvolvimento  Cient\'{\i}fico e Tecn\'ogico - CNPq, grant nos. 
312369/2019-0 (S.S.P.) and 309262/2019-4  (G.K.), Funda\c{c}\~ao de 
Amparo \`a Pesquisa do Estado de S\~ao Paulo - FAPESP, grant no.  2018/25225-9 
(S.S.P. and G.K.).

 \bibliographystyle{SciPost_bibstyle} 
\bibliography{SciPost_Example_BiBTeX_File.bib}

\begin{thebibliography}{1}
\providecommand{\url}[1]{\texttt{#1}}
\providecommand{\urlprefix}{URL }
\expandafter\ifx\csname urlstyle\endcsname\relax
  \providecommand{\doi}[1]{doi:\discretionary{}{}{}#1}\else
  \providecommand{\doi}{doi:\discretionary{}{}{}\begingroup
  \urlstyle{rm}\Url}\fi
\providecommand{\eprint}[2][]{\url{#2}}

\bibitem{brown1954}
R.~Hanbury-Brown and R.~Q. Twiss,
\newblock \emph{{A New type of interferometer for use in radio astronomy}},
\newblock The London, Edinburgh, and Dublin Philosophical Magazine and Journal
  of Science \textbf{45}(366), 663 (1954),
\newblock \doi{10.1080/14786440708520475}.

\bibitem{GGLP60}
G.~S. L.~W. Goldhaber, G. and A.~Pais,
\newblock \emph{{Influence of Bose-Einstein Statistics on the Antiproton-Proton
  Annihilation Process}},
\newblock Physical Review \textbf{120}(1), 300 (1960),
\newblock \doi{10.1103/PhysRev.120.300}.

\bibitem{lisa2005}
M.~A. Lisa, S.~Pratt, R.~Soltz and U.~Wiedemann,
\newblock \emph{{FEMTOSCOPY IN RELATIVISTIC HEAVY ION COLLISIONS: Two Decades
  of Progress}},
\newblock Annual Review of Nuclear and Particle Science \textbf{55}(1), 357
  (2005),
\newblock \doi{10.1146/annurev.nucl.55.090704.151533}.

\bibitem{Lednicky:1981su}
R.~Lednicky and V.~L. Lyuboshits,
\newblock \emph{Final state interaction effect on pairing correlations between
  particles with small relative momenta},
\newblock Yad. Fiz. \textbf{35}, 1316 (1981).

\bibitem{Hadjimichef:1998rx}
D.~Hadjimichef, G.~Krein, S.~Szpigel and J.~S. da~Veiga,
\newblock \emph{{Mapping of composite hadrons into elementary hadrons and
  effective hadronic Hamiltonians}},
\newblock Annals Phys. \textbf{268}, 105 (1998),
\newblock \doi{10.1006/aphy.1998.5825},
\newblock \eprint{hep-ph/9805459}.

\end{thebibliography}

\nolinenumbers

\end{document}